\documentclass[twocolumn,showpacs,prl,10pt,superscriptaddress]{revtex4}
\usepackage[english]{babel}
\usepackage{amsmath,amssymb}
\usepackage{times}
\usepackage{epsfig}

\def\tcw{69.5mm}
\def\vc#1{\mathbf{#1}}
\def\avg#1{{\langle{#1}\rangle}}
\def\RH{R_{\text{H}}}
\def\rH{r_{\text{H}}}
\def\Re{\operatorname{Re}}

\begin{document}
\title{Temperature Dependence of Hall Response in Doped
Antiferromagnets}
\author{Darko \surname{Veberi\v c}}
\affiliation{J.\ Stefan Institute, SI-1001 Ljubljana, Slovenia}
\affiliation{Nova Gorica Polytechnic, SI-5001 Nova Gorica, Slovenia}
\author{Peter \surname{Prelov\v sek}}
\affiliation{J.\ Stefan Institute, SI-1001 Ljubljana, Slovenia}
\affiliation{Faculty of Mathematics and Physics, University of Ljubljana,
SI-1001 Ljubljana, Slovenia}
\date{\today}

\begin{abstract}
Using finite-temperature Lanczos method the frequency-dependent Hall
response is calculated numerically for the $t$-$J$ model on the square
lattice and on ladders. At low doping, both the high-frequency $\RH^*$
and the d.c.\ Hall coefficient $\RH^0$ follow qualitatively similar
behavior at higher temperatures: being hole-like for
$T>T_s!\!\approx\!!1.5J$ and weakly electron-like for $T<T_s$.
Consistent with experiments on cuprates, $\RH$ changes, in contrast to
$\RH^*$, again to the hole-like sign below the pseudogap temperature
$T^*$, revealing a strong temperature variation for $T\to0$.
\end{abstract}
\maketitle

The anomalous behavior of the Hall constant $\RH$ in the normal state
of cuprates \cite{ong} remains the challenge for theoreticians for
over a decade. Two aspects, possibly interrelated, are evident and
should be understood: a) the d.c.\ $\RH^0$ at low temperatures $T\to0$
is clearly doping dependent. In the prototype material
La$_{2-x}$Sr$_x$CuO$_4$ (LSCO) it changes from positive
$\RH^0\propto1/x$ at low doping $x<x^*\!\!\approx\!\!0.3$, consistent
with the picture of hole-doped (Mott-Hubbard) insulator, to the
electron-like $\RH^0<0$ at $x>x^*$ in agreement with the usual band
picture. b) $\RH^0$ is also strongly temperature dependent, both at
low doping and optimum doping. At optimum doping, the attention has
been devoted to the anomalous variation of the Hall angle
$\theta_{\text{H}}\propto T^2$ in YBa$_2$Cu$_3$O$_7$ \cite{ande}. On
the other hand, at low hole concentration $c_h<0.15$, $\RH(T)$ in LSCO
has been shown to follow an universally behaved \cite{hwan} decrease
with $T$ in which $\RH^0(T\!\!\to\!\!0)$ and the characteristic
temperature $T^*$ of vanishing $\RH^(T^*)\sim 0$ both scale with
$c_h$. In underdoped cuprates, the same $T^*(c_h)$ has been in fact
associated with the (large) pseudogap crossover scale in uniform
susceptibility $\chi_0(T)$, in-plane resistivity $\rho(T)$, specific
heat $c_v(T)$, and some other quantities \cite{imad}.

A number of theoretical investigations have addressed the first
question, i.e.\ the doping dependence of $\RH$ in models of strongly
correlated electrons, in particular within the $t$-$J$ model and the
Hubbard model on a planar lattice. The advantage is that one can study
the dynamical Hall response and the d.c.\ Hall constant as a ground
state ($T=0$) property, in particular in systems with finite
transverse dimension \cite{cast,rojo} and in the ladder geometry
\cite{tsun,prel}. It has been also shown that within the $t$-$J$
model the change from a hole-like to an electron-like Hall response
can be qualitatively reproduced by studying the high-frequency
$\RH^*=\RH(\omega\!\!\to\!\!\infty)$ \cite{shas}, analytically
tractable at $T\to\infty$. Recently, a connection of the reactive
$\RH^0(T\!\!=\!\!0)$ to the charge stiffness has also been found
\cite{zoto}.

The anomalous temperature dependence of $\RH(T)$, being the main
subject of this work, has been much less clarified in the literature,
The Hall mobility $\mu_{\text{H}}(T)$ of a single charge carrier in
the Mott-Hubbard insulator has been first evaluated within the
generalized retraceable path approximation \cite{brin}. The
high-frequency $\RH^*(T)$ has been calculated using the high-$T$
expansion \cite{shas}. At low doping, $c_h<0.15$, it has been observed
that on decreasing temperature $\RH^*$ is also decreasing instead of
approaching presumed (larger) semiclassical and experimentally
observed d.c.\ result $\RH^c=1/c_he_0\approx4\RH^*(T\!\!=\!\!\infty)$.
Related are the conclusions of the quantum Monte-Carlo study of the
planar Hubbard model \cite{assa}, where close to the half-filling
electron-like $\RH^*<0$ has been found at low $T$. The same has been
claimed generally for $\RH(\omega)$ even for low $\omega$ \cite{assa}.
Quite controversial are also results for $\RH^0(T)$ on ladders
\cite{tsun}. In regard to that, we should also mention the
questionable relation of the off-diagonal $\sigma_{xy}$ to the orbital
susceptibility $\chi_d$ \cite{wido,rojo}, potentially useful as an
alternative route to the understanding of $\RH^0(T)$ \cite{vebe}.

In the following we present numerical results for the dynamical
$\RH(\omega)$, as obtained within the low doping regime of the $t$-$J$
model using the finite-temperature Lanczos method (FTLM)
\cite{jplan,jprev}. The aim of this letter is to approach the
low-$\omega$ and low-$T$ limit as much as possible and to investigate
the relation between $\RH^*(T)$ and $\RH^0(T)$. We find these two
quantities essentially different for $T<T^*$, establishing the
pseudogap scale $T^*<J$ both in the ladder and planar systems.

We study the $t$-$J$ model in an external homogeneous magnetic field
$\vc{B}=\operatorname{curl}\vc{A}$,
\begin{align}
H(\vc{A})=&-t\sum_{\avg{ij}s}(
\mathrm{e}^{\mathrm{i}\theta_{ij}}\tilde{c}_{is}^\dagger
\tilde{c}_{js}^{\phantom{\dagger}}+\text{H.c.})+
\nonumber
\\
&+J\sum_{\avg{ij}}(\vc{S}_i\!\cdot\!\vc{S}_j-\tfrac14n_in_j),
\label{eq1}
\end{align}
where the (inhomogeneous) vector potential enters the phases
$\theta_{ij}=e\vc{A}(\vc{r}_i)\!\cdot\!\vc{r}_{ij}$. The hopping is
only between the nearest neighbors $\avg{ij}$. Projected fermionic
operators $\smash{\tilde{c}^{\phantom{\dagger}}_{is}}$,
$\smash{\tilde{c}^\dagger_{is}}$ do not allow for the double occupancy
of sites.

In order to calculate the dynamical Hall coefficient
\begin{equation}
\RH(\omega)=\frac{\partial\rho_{xy}(\omega)}{\partial B}\Big|_{B\to0}=
\frac{\sigma_{xy}(\omega)}{B\sigma_{xx}(\omega)\sigma_{yy}(\omega)}\Big|_{B\to0},
\label{eq2}
\end{equation}
the conductivity tensor is evaluated within the linear response
theory,
\begin{equation}
\sigma_{\alpha\beta}(\omega)=\frac{\mathrm{i}e^2}{N\omega^+}
\left[\avg{\tau_{\alpha\beta}}-\mathrm{i}\!\int_0^\infty\!\!\!\!
\text{d}t\,\mathrm{e}^{\mathrm{i}\omega t}\avg{[j_\alpha(t),j_\beta]}\right],
\label{eq3}
\end{equation}
where in the presence of $B\neq0$ the particle current $\vc{j}$ and the
stress tensor $\underline{\tau}$ operators are given by
\begin{align}
\vc{j}&=t\sum_{\avg{ij}s}\vc{r}_{ij}(
\mathrm{i}\mathrm{e}^{\mathrm{i}\theta_{ij}}
\tilde{c}_{is}^\dagger\tilde{c}_{js}^{\phantom{\dagger}}+
\text{H.c.}),
\nonumber
\\
\underline{\tau}&=t\sum_{\avg{ij}s}\vc{r}_{ij}\otimes\vc{r}_{ij}
(\mathrm{e}^{\mathrm{i}\theta_{ij}}
\tilde{c}_{is}^\dagger\tilde{c}_{js}^{\phantom{\dagger}}+
\text{H.c.}).
\label{eq4}
\end{align}

On a square lattice with $N$ sites and periodic boundary conditions
(b.c.) one cannot apply arbitrary magnetic field $B$ since only
quantized $B=B_m=m\Phi_0 a^2/N$ can be made compatible with the
periodic b.c.\ \cite{assa}. Therefore the smallest but finite $B=B_1$
is used in calculations. The square lattices used are in general
Euclidean (tilted) $N=l^2+n^2$, in particular we investigate systems
$N=10$, 16, 18. On the other hand, the ladder geometry of $N=L\times
M$ sites with the periodic b.c.\ in the $L$ direction and open b.c.\
in the perpendicular $M$ direction allows for any finite $B\neq0$, the
fact already used in several $T=0$ calculations \cite{cast,rojo,prel}.
The advantage of ladder systems is also the existence of the reference
ground-state results $\RH^0(T\!\!=\!\!0)$ which seem to be better
understood \cite{prel,zoto}. Furthermore, at low doping they reproduce
the simple semiclassical behavior
$\RH^0(T\!\!=\!\!0)\sim\RH^{\text{c}}=1/c_he_0$.

Dynamical components $\sigma_{\alpha\beta}(\omega)$ are evaluated
using the FTLM \cite{jplan}, employed so far for various dynamic and
static quantities within the $t$-$J$ model \cite{jprev}, among them
also the $B=0$ optical conductivity
$\sigma(\omega)=\sigma_{\alpha\alpha}(\omega)$ on a square
lattice. Comparing to the diagonal $\sigma_{\alpha\alpha}$, the
evaluation of the off-diagonal $\sigma_{xy}(\omega)$ is more demanding
for several reasons: a) the introduction of $B>0$ in the model
\eqref{eq1} breaks the translational invariance and prevents the
reduction of the basis states in the Lanczos procedure, hence
available finite-size systems are somewhat smaller, b) we expect
$\sigma_{xy}(\omega)\propto B$ while $\sigma_{xy}(B\!\!=\!\!0,\omega)$
does not vanish identically within the FTLM; consequently larger
sampling over initial wavefunctions \cite{jplan,jprev} are needed to
reduce the statistical error, c) on a finite square lattice the
reference result $\RH^0(T\!\!=\!\!0)$ is not meaningful for $B_m>0$,
while in ladder systems it is quite sensitive to the introduction of
an additional flux \cite{prel}. Nevertheless, in general, restrictions
for the validity of the FTLM results are similar to other quantities.
Through the thermodynamic partition function $Z(T_{\text{fs}})=Z^*$,
we can define the marginal finite-size $T_{\text{fs}}$ below which too
few levels contribute to the average and results loose the
thermodynamic validity \cite{jprev}. In the following, we analyze
results for $J=0.4t$ at low hole doping $c_h=N_h/N$ ($N_h=1$, 2). In
this regime we can estimate $T_{\text{fs}}/t \sim 0.15-0.2 \lesssim
0.5J/t$.

Let us first present results for the dynamical $\RH(\omega)$. In
Fig.~1 we show the normalized real part $\rH=e_0c_h\Re\RH$ for systems
with a single hole $N_h=1$. In the evaluation of $\RH(\omega)$ from
Eq.~\eqref{eq2} we insert complex $\sigma_{\alpha\alpha}$ at $B=0$ and
the most sensitive quantity remains $\sigma_{xy}(\omega)$ calculated
at $B=B_1$ on a square lattice and $B\sim0.3B_1$ on ladders. In the
presentation of results an additional frequency smoothening
$\delta=0.2t$ is used. The normalization of $\RH$ is chosen such that
at low doping $\rH=1$ would show up in the case of the semiclassical
result.

\begin{figure}[tb]
\centering
\epsfig{file=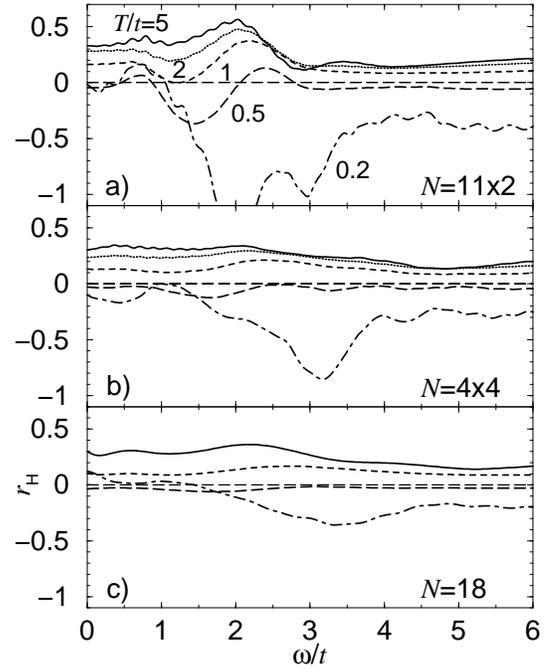,width=\tcw}
\caption{Dynamical Hall response $\rH(\omega)=e_0c_h \Re\RH(\omega)$
for different temperatures $T/t$ and various systems with a single
hole $N_h=1$: a) 2-leg ladder with $L=11$, b) 4-leg ladder with
$L=4$, and c) square lattice with $N=18$ sites.}
\label{fig1}
\end{figure}

In Fig.~1 several common features of $\RH$ in the ladder geometry and
in the 2D systems are recognized:

\noindent a) $\rH(\omega)$ is quite smoothly varying function
of $\omega$, at least in contrast to strongly $\omega$-dependent
$\Re\sigma(\omega)$ on a 2D system, which is found \cite{jprev} to
decay with an anomalous relaxation rate $1/\tau(\omega) \propto
\omega+\xi T$.

\noindent b) At high temperatures $T>t$ we get a hole-like
$\rH>0$ for all systems. In this regime $\rH(\omega)$ is very
smooth, in particular for the $M=4$ ladder and the 2D lattice. 

\noindent c) For low temperatures $T<t$, $\rH(\omega)$ is less smooth
and the dependence is more pronounced for the 2-leg ladder. On the
other hand, $M=4$ ladder clearly approaches the behavior of the 2D
system, whereby both of the latter show quite a modest variation of
$\rH(\omega)$. In all systems the resonances (and the variation)
visible in $\rH(\omega)$ at high $\omega>t$ reflect the predominantly
local physics of the hole motion and are thus not related to a current
relaxation rate deduced from $\sigma(\omega)$.

Results for $\rH(\omega)$ are the basis for the calculation of
high-frequency $\rH^*=\rH(\omega\!\!=\!\!\infty)$ as well as the d.c.\
limit $\rH^0=\rH(\omega\!\!\to\!\!0)$. The latter is more sensitive
since in a finite system (even at $T>0$)
$\sigma_{\alpha\beta}(\omega\!\!\to\!\!0)$ can be singular due to the
coherent charge transport in a system with periodic b.c.. The coherent
transport shows up in a finite (but small) charge stiffness
\cite{jprev}, which should be omitted in the evaluation of
Eq.~\eqref{eq2}. In any case, one should take into account proper
$\omega\to0$ behavior of dissipative systems at $T>0$ which is
different in ladders and in 2D lattices, respectively: a) On a ladder
we get in the leading order of $\omega\to0$ a normal conductance along
the $x$-direction, i.e.\
$\sigma_{xx}(\omega\!\!\to\!\!0)\sim\sigma_0$, but a finite
polarizability along the $y$-direction,
$\sigma_{yy}(\omega\!\!\to\!\!0)\propto\omega\chi_{yy}^0$. Hence, we
expect $\sigma_{xy}\propto\omega$ and finite $\rH^0$. b) For a
macroscopic isotropic 2D system we get
$\sigma_{\alpha\alpha}(\omega\!\!\to\!\!0)\to\sigma_0$ and we expect
as well $\sigma_{xy}\to\sigma^0_{xy}$, leading to finite $\rH^0$.

\begin{figure}[tb]
\centering
\epsfig{file=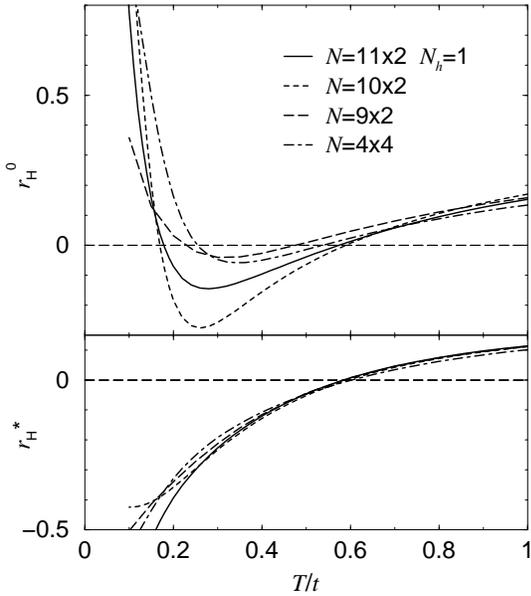,width=\tcw}
\caption{D.c.\ Hall constant $\rH^0$ and the infinite-frequency
$\rH^*$ vs.\ $T/t$ for various ladders $L\times M$ with a single hole
$N_h=1$.}
\label{fig2}
\end{figure}

\begin{figure}[tb]
\centering
\epsfig{file=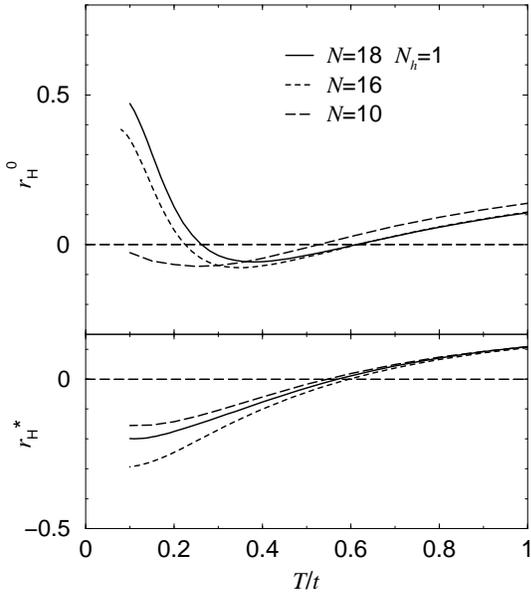,width=\tcw}
\caption{$\rH^0$ and $\rH^*$ vs.\ $T/t$ for different square lattices
with $N$ sites and a single hole $N_h=1$.}
\label{fig3}
\end{figure}

\begin{figure}[tb]
\centering
\epsfig{file=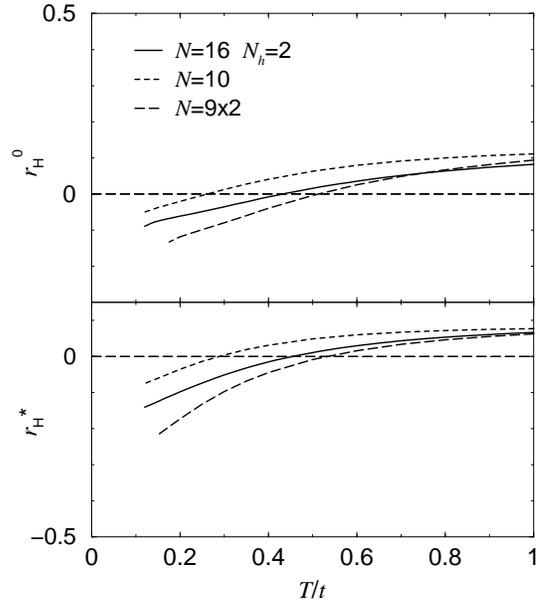,width=\tcw}
\caption{$\rH^0$ and $\rH^*$ vs.\ $T/t$ for different square and ladder
lattices with $N$ sites and $N_h=2$.}
\label{fig4}
\end{figure}

In Fig.~2 we present results for $\rH^*(T)$ and $\rH^0(T)$ for the
ladder systems with $N_h=1$. Results are shown for 2-leg ladders with
various lengths $L=9$, 10, 11 and for $M=4$ ladder with $L=4$. Since
$\rH^0$ and $\rH^*$ are properly scaled, for given $M$ curves are
expected to approach a well defined macroscopic limit at $L\to\infty$.
In fact, $\rH^*$ are nearly independent of $L$ (as well as of $M$)
down to $T\sim T_{\text{fs}}$. A crossover at $T_s\sim0.6t$ from a
hole-like $\rH^*>0$ into a electron-like $\rH^*<0$ can be explicitly
observed. $\rH^0$ results are more size ($L$) dependent, nevertheless
they reveal a crossover nearly at the same $T\sim T_s$. In contrast to
$\rH^*$ which remains negative for the whole regime $T<T_s$, $\rH^0$
changes sign again at $T\!\!=\!\!T^*\sim0.2t$. Although our data for
$T^*$ are more scattered the crossover into the hole-like
$\rH^0(T\!\!<\!\!T^*)>0$ is expected. Namely, from the ground state
calculations in same systems \cite{prel} we know that
$\rH^0(T\!\!=\!\!0)\sim1.5$ and $\rH^0(T\!\!=\!\!0)\sim1.2$ for $M=2$
and $M=4$ ladders, respectively. Therefore, it is not surprising that
the observed dependence $\rH^0(T\!\!<\!\!T^*)$ is very steep.

Corresponding results for the planar lattice in Fig.~3 are both
qualitatively and quantitatively similar. Note, that at low doping the
limiting value $\rH^*(T\!\!\to\!\!\infty)=1/4$ agrees with the
analytical result \cite{shas}, while obtained
$\rH^0(T\!\!\to\!\!\infty)\sim0.3$ is also quite close. Again, the
crossover into an electron-like regime appears at $T_s\sim0.6t$. For
larger sizes $N\geq16$ the lower crossover $T^*\sim0.2t$ is visible as
well. In finite 2D systems a reference numerical result at $T=0$ does
not exist, however, the analytical theory \cite{prel1} indicates that
in a macroscopic limit with a single hole ($N_h=1$) in an ordered
antiferromagnet one should get $\rH^0=1$.

In numerically available systems, $N_h=2$ represents already a
substantial doping. Therefore, results for $\rH^0$ and $\rH^*$ shown
in Fig.~3 should be interpreted in relation with the corresponding
finite doping $c_h$. Main message of Fig.~3 is that upper crossover
$T_s$, still nearly the same in both $\rH^0(T)$ and $\rH^*(T)$, shifts
down quite systematically with increasing $c_h$, i.e.\ with decreasing
size $N$ at given $N_h$. At least in ladder systems at $c_h<0.3$, we
still find $\rH^0(T\!\!=\!\!0)>0$ in the ground state \cite{prel},
therefore also the lower crossover $T^*<T_s$ is expected. However, we
cannot detect such a crossover in $\rH^0(T)$ down to
$T_{\text{fs}}\sim0.15t$, not surprisingly since also the experimental
value, e.g.\ in LSCO at $c_h>0.1$, is $T^*<600\,\text{K}\sim0.15t$
(assuming $t\sim0.4\,e$V).

Let us finally comment on the relation of the d.c.\ $\sigma_{xy}^0$ to
the orbital susceptibility $\chi_d$ in a macroscopic 2D
system. Namely, $\tilde{\sigma}_{xy}^0=eB\partial\chi_d/\partial\mu=
eB(\partial\chi_d/\partial c_h)(\partial c_h/\partial\mu)$, (where
$\mu$ denotes chemical potential) was derived using seemingly quite
general thermodynamic relations \cite{wido,rojo}, but at the same time
put under question \cite{rojo}. Since the d.c.\
$\sigma^0_{\alpha\alpha}(T)>0$ is quite a smooth function the above
relation seems to yield also a qualitative connection between
$\chi_d(T)$ and $\RH^0(T)$. The situation should be particularly
simple at low doping (but not too low $T$), where $\partial
c_h/\partial\mu\sim c_h/T$ and $\chi_d\propto c_h$ is expected, and
consequently $\tilde{\sigma}_{xy}\propto -B\chi_d/T$. Indeed, results
for $N_h=1$ indicate \cite{vebe} that both crossovers $T_s$ and $T^*$
appear also as a change of sign in $\chi_d(T)$ nearly at the same
values. Here, the intermediate regime $T^*<T<T_s$ corresponds to an
anomalous paramagnetic response $\chi_d>0$. On the other hand, it is
quite evident from our results that the relation is not valid at high
$T\gg t$. Namely, in this regime $\sigma^0_{\alpha\alpha}\propto1/T$
and $\sigma_{xy}^0\propto B/T^2$ \cite{brin} is obtained, leading to
$\RH^0(T\!\!\to\!\!\infty) \sim\text{const}$. On the other hand, from
the high-$T$ expansion $\chi_d\propto1/T^3$ is acquired \cite{vebe},
so that the assumed relation would demand $\tilde\sigma_{xy}^0\propto
B/T^4$, in conflict with previous $\sigma_{xy}^0\propto B/T^2$.

In conclusion, we have presented results for both dynamical and d.c.\
Hall constant within the $t$-$J$ model on ladders and on square
lattices. The main novel point is the observation of two crossover
temperatures $T_s$ and $T^*$ which are at low doping generally present
in all systems. Both $\RH^*$ and $\RH^0$ are positive at $T>T_s$ and
change sign at $T_s$. While $\RH^*(T\!\!<\!\!T_s)$ stays negative,
$\RH^0$ reveals a sign change into a hole-like behavior at
$T\!\!=\!\!T^*<T_s$ as well as steep variation of
$\RH^0(T\!\!<\!\!T^*)$. This reconciles some seemingly controversial
theoretical results \cite{shas,assa}. Our results are in agreement
with high-$T$ expansion results for $\RH^*(T)$ which at low $c_h$ also
show decreasing positive values with decreasing $T$. Quantum Monte
Carlo results within the Hubbard model for $\RH(\mathrm{i}\omega)$
correspond effectively to high (imaginary) frequencies and low $T$,
and being negative they are in agreement with our findings for
$\RH^*$.

How should we understand the above numerical results? At high $T\gg t$
and low doping $c_h\ll1$, $\RH^*$ as well as $\RH^0$ are governed by a
loop motion (that is where the dependence on $B\neq0$ comes from) of a
hole within a single plaquette \cite{brin,shas}. One expects
$\RH^*>0$, but $\rH^*=1/4$ is a non-universal value which e.g.\
depends on the lattice coordination \cite{shas}. The electron-like
$\rH^*(T\!\!=\!\!0)<0$ represents an instantaneous Hall response
within the ground state near half filling is harder to explain, but is
clearly the signature of strong correlations. On the other hand at low
$T$, $\RH^0$ tests the (low energy) quasiparticle
properties. Evidently, at low doping and $T<T^*$ at least a single
hole in an antiferromagnetic spin background behaves as a well defined
hole-like quasiparticle leading to $\rH^0(T\!\!\to\!\!0)\sim1$ both in
2D \cite{prel1} and in ladders \cite{prel}. Our results for vanishing
$\RH^0(T^*)\sim 0$ indicate that the quasiparticle character is
essentially lost at quite low $T\sim T^*<J$, with the pseudogap scale
$T^*(c_h)$ decreasing with doping. Such phenomenon is possibly
consistent with the scenario of electrons being effectively composite
particles (spinons and holons) in strongly correlated systems
\cite{ande,laug}, at least at $T>T^*(c_h)$, whereby $T^*(c_h)$
vanishes at optimum doping.

Finally, let us note that our results for $\RH^0$ are in several
aspects consistent with experiments on cuprates, and with LSCO in
particular. At low doping $c_h<0.1$ we find $T^*\sim J/2$, close to
the observed $T^*\sim 800\,$K. At the same time, we find a very steep
dependence in $\RH^0(T\!\!<\!\!T^*)$. With increasing $c_h$,
$T^*(c_h)$ seems to have desired decreasing tendency, although to
establish this beyond a reasonable doubt more work is needed.

Authors acknowledge the support of the Ministry of Education, Science
and Sport of Slovenia.

\end{document}